\documentclass[aps, pre, reprint, amsmath, amssymb, superscriptaddress]{revtex4-1}
\usepackage{bm}
\usepackage{graphicx}
\usepackage{transparent}
\usepackage{tikz}
\usepackage{placeins}
\usepackage{mathrsfs}
\usepackage{wasysym}
\usepackage{physics}
\usepackage{nicefrac}
\usepackage{color}
\usepackage{comment}
\usepackage{soul}
\usepackage{times}
\usepackage{txfonts}
\usepackage[normalem]{ulem}
\usepackage[colorlinks = true, linkcolor = blue, urlcolor = blue, citecolor = blue, pdfusetitle]{hyperref}

\begin{document}

\title{Universal splitting of nonequilibrium phase transitions in driven Potts heat engines}
\author{Vitória T. Henkes}
\author{Gustavo A. L. Forão}
\affiliation{Universidade de São Paulo,
Instituto de Física,
Rua do Matão, 1371, 05508-090
São Paulo, SP, Brazil}
\author{Andre C. Barato}
\affiliation{Department of Physics, 
University of Houston, 
Houston, Texas 77204, USA}
\author{Carlos E. Fiore}
\affiliation{Universidade de São Paulo,
Instituto de Física,
Rua do Matão, 1371, 05508-090
São Paulo, SP, Brazil}

\date{\today}

\begin{abstract} We investigate nonequilibrium phase transitions and thermodynamic properties of driven Potts models coupled to two thermal reservoirs at different temperatures. 
The interplay with the multi-state structure of the Potts model leads to distinct phase-transition scenarios. The proposed driving scheme breaks the symmetry between the Potts
states and gives rise to multiple transitions within the ordered phases, where the phases are characterized by different numbers of stable fixed points. This number can vary from $q$ fixed points to $1$ fixed point,
corresponding to $q-1$ distinct transitions. The driving scheme strongly affects the thermodynamic operation regimes, allowing the system to operate as a heat engine.
We demonstrate that  these phenomena persist for both a mean-field (MF) model and a two-dimensional model, confirming the robustness of the results with respect 
to dimensionality. For the MF case, we also develop a phenomenological description in the strongly ordered regime that yields analytical expressions for the power, heat currents, and efficiency. 
Our results show how multi-state collective interactions, nonequilibrium driving, and thermal bias jointly generate a rich phase structure and the emergence of a collective heat engine.

\end{abstract}

\maketitle

\section{Introduction}

While equilibrium statistical mechanics and allied thermodynamics provide a robust framework for describing macroscopic properties and phase transitions in generic systems at thermal equilibrium \cite{salinas2001introduction, yeomans1992statistical}, stochastic thermodynamics extends this framework to nonequilibrium systems, where fluctuations play a central role and can be described by a Markovian dynamics. The assumption of local detailed balance provides a consistent connection between stochastic dynamics and thermodynamic quantities, leading to fundamental relations such as Jarzynski \cite{jarz} and Crooks fluctuation theorems \cite{crooks1999entropy} and thermodynamic uncertainty relations (TURs) \cite{barato2015thermodynamic, barato2018bounds}. 

Nonequilibrium phase transitions arise in a wide variety of contexts, ranging from absorbing-state \cite{marro2005} and voter-model dynamics \cite{harunari2017partial, encinas2018fundamental, encinas2019majority, fiore23, fiore25} to social \cite{castellano2009statistical}, biological \cite{rapoport1970sodium, lynn2021broken}, and competition dynamics at different temperatures \cite{Garrido1987, Garrido1989, Blote1990,Tome_1991, Gambetta2019, aguilera2023nonequilibrium, Yan2023, dutta2025, seifert2012stochastic, mamede2023, forao2024splitting,forao2023powerful}. They may involve continuous or discontinuous changes in collective behavior, with their critical properties depending on the underlying dynamics, symmetries, and dimensionality. Among the different models exhibiting collective behavior and phase transitions, Ising and Potts models \cite{RevModPhys.54.235} constitute paradigmatic examples due to their simplicity and broad range of applications. They have been extensively studied at equilibrium and, more recently, extended to the nonequilibrium context, leading to important differences, such as the existence of singular behavior in the entropy production \cite{martynec2020entropy, fiore21, herpich, herpich2, noa2019entropy, zhang2016critical, tome2012}.

Recently, a novel class of nonequilibrium phase transitions \cite{forao2024splitting} has been uncovered in driven Ising models, exhibiting behavior that differs significantly from that of conventional nonequilibrium phase transitions \cite{marro2005}. They are also different from phase transitions in driven Ising models at a single temperature \cite{herpich, herpich2, barato2016cost} and in undriven systems at different temperatures \cite{fiore21, mamede2025collectiveheatenginesdifferent}. Such distinctive behavior emerges from the intricate interplay between biased forces and multiple thermal reservoirs \cite{forao2024splitting, forao2025universal}, with the nature of the transition crucially depending on the initial ordered state: it is continuous when starting from the ``down"-spin phase and discontinuous when starting from the ``up"-spin phase. We call this phenomenon the splitting of phase transitions. It reveals that a single system can exhibit qualitatively distinct phase-transition scenarios under identical control parameters. Furthermore, the critical behavior deviates from that of standard spontaneous symmetry-breaking transitions, exhibiting a distinct set of critical exponents.

We consolidate this novel class of phase transitions by introducing a  driven Potts model in contact with different temperatures.  
This driven Potts model displays two main phenomena. First, there is splitting of the ordered phase into $q$ different phases, where $q$
is the number of states of a single unit. These phases are characterized by the number of stable fixed points, which can be any number between $1$ and $q$.
Hence, there are $q$ phases and $q-1$ transitions within the ordered phase. Second, there is a regime that allows for the emergence of a heat 
engine within the phase with $q$ stable fixed points and for initial conditions that go to one of the fixed points. 

Our results are obtained for both a MF model and a two-dimensional model. The MF model allows for semi-analytical calculations. For the 
two-dimensional model we resort to numerical simulations. Despite the absence of closed-form expressions for the order parameter and thermodynamic quantities for the MF model, 
we develop a comprehensive phenomenological description for generic $q$ states. This framework successfully captures all optimal performance 
regimes and excellently matches our semi-analytical results for the MF model.

This paper is structured as follows. In Sec.~\ref{model}, we introduce the model, the dynamics, and define the thermodynamic quantities. 
Sec.~\ref{MF-results1} contains the splitting of the ordered phase for the MF model, while Sec.~\ref{MF-results2} contains the emergence 
of a collective heat engine in our model. The results for the two-dimensional model that displays the same two phenomena
are shown in \ref{SL-results}. We conclude in Sec.~\ref{conclusion}. 

\section{General model} \label{model}

\subsection{System, driving, and transition rates}

We consider a system composed of $N$ interacting units $i \in \{1,2,\dots,N\}$, each occupying one of $q\geq 3$ discrete states $s_i \in \{0,1,\dots,q-1\}$. The case $q = 2$, which corresponds to the Ising model, has been previously considered for autonomous heat engines \cite{forao2023powerful, mamede2023, gatien}. A microstate is specified by the configuration $s\equiv(s_1,\dots,s_i,\dots,s_N)$, which has the energy
\begin{equation}
E(s)\equiv\epsilon\sum_{i=1}^N\sum_{\delta=1}^{k/2}\delta_{s_i,s_{i+\delta}},
\label{eq1}
\end{equation}
where $\epsilon<0$ sets the interaction strength, and each unit interacts with its $k$ nearest neighbors. 

\begin{figure}
    \centering
    \includegraphics[width=1.\linewidth]{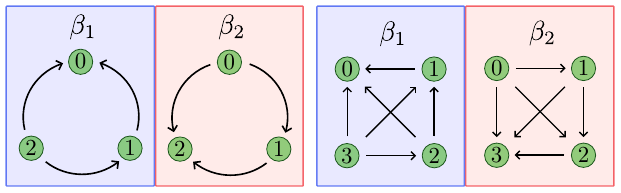}
    \caption{Schematics of the driving forces. Small green circles denote the unit states $s_i$, and arrows indicate the favored direction, i.e. $d^{(\nu)}_{s's}>0$, of the driving force for $q=3$ (left) and $q=4$ (right). The driving favors higher (lower) states due to the contact with the hot (cold) thermal reservoir.}
    \label{fig:schematics2}
\end{figure}

The system is simultaneously coupled to two thermal reservoirs $\nu \in \{1,2\}$ at inverse temperatures $\beta_1>\beta_2$, where $\nu=1$ ($\nu=2$) 
denotes the cold (hot) reservoir. The parameter $\nu$ represents different transition pathways between a pair of states. Our model also incorporates a driving force $F$ related to transition between states $s$ and $s'$ through  the quantity 
\begin{equation}
F^{(\nu)}_{s's}\equiv Fd^{(\nu)}_{s's},
\end{equation}
where $s'=(s_1,\dots,{\tilde s}_i,\dots,s_N)$, with ${\tilde s_i}\neq s_i$. The increments $d^{(\nu)}_{s's}$ satisfy the antisymmetry relation $d^{(\nu)}_{s's} = -d^{(\nu)}_{ss'}$. We further assume that they depend on the reservoir according to $d^{(1)}_{s's} = -d^{(2)}_{s's}$, corresponding to opposite directions of the driving force for the two reservoirs. We investigate the influence of the driving force through the scheme  
illustrated in Fig.~\ref{fig:schematics2}. 

The increments of the driving force $F$ have the values $\pm 1/(q-2)$ and are distributed along the cycle $0\rightarrow 1\rightarrow 2,\dots,\rightarrow q-1 \rightarrow 0$ in the following way. The increment
$d^{(1)}_{s's}$ ($d^{(2)}_{s's}$) is negative (positive) for $s'>s$ and positive (negative) for $s'<s$. 
Using these increments, the full cycle affinity of the cycle $0\rightarrow 1\rightarrow 2,\dots,\rightarrow q-1 \rightarrow 0$ is $\beta_2F$ for $\nu=2$ and $-\beta_1F$ for $\nu=1$, independent of  the value of $q$. 
We point out that other transitions are also allowed, which means that the system can go through many other cycles. As an example, for $q=4$, an unit can go through the cycle 
$0\rightarrow 1\rightarrow 3\rightarrow 0$, which has affinities $\beta_2F/2$ for $\nu=2$ and $-\beta_1F/2$ for $\nu=1$.

Transition rates $\omega^{(\nu)}_{s's}$ for a transition $s\rightarrow s'$ are given by
\begin{eqnarray}
    \omega^{(\nu)}_{s's}&=&\Gamma\exp\left[-\frac{\beta_\nu}{2}Q^{(\nu)}_{s's}\right],
    \label{transition}
\end{eqnarray}
where  
\begin{equation}
Q^{(\nu)}_{s's}\equiv \Delta E_{s's}-F^{(\nu)}_{s's}
\label{heat1}
\end{equation}
is the heat absorbed from reservoir $\nu$ due to the transition $s\rightarrow s'$ and $\Delta E_{s's}\equiv E(s')-E(s)$.
The parameter $\Gamma$ sets the time scale of the transitions. For nonzero transition rates, the states $s$ and $s'$ differ by a flip of a single unit.

\subsection{Mean-field model}

In the mean-field model, all units interact with each other. The $q^N$ microstates $s$ can be fully specified by the occupation numbers $\mathbf{N}\equiv\{N_0,N_1,\dots,N_{q-1}\}$, where $N_\alpha$ counts the units in state $\alpha\in\{0,...,q-1\}$ and $\sum_\alpha N_\alpha = N$. 
In terms of these variables, Eq.~(\ref{eq1}) is  rewritten as
\begin{equation}
E(\mathbf{N})=\frac{\epsilon}{2N}\sum_{\alpha}N_\alpha(N_\alpha - 1), 
\end{equation} 
where the extra factor of $1/N$ ensures an extensive energy. Defining the states $\mathbf{N}_{\alpha} \equiv \{\ldots, N_{\alpha'} + 1, \ldots,
N_\alpha - 1, \ldots\}$ and $\mathbf{N}_{\alpha'} \equiv \{\ldots, N_{\alpha'} - 1, \ldots,N_\alpha + 1, \ldots\}$, the master equation for the mean-field model in terms of the occupation numbers 
is written as 
\begin{equation}
{\dot p}_{\mathbf{N}_{\alpha}}(t) = \sum_{\nu=1}^{2} \sum_{\mathbf{N}_{\alpha}'}J^{(\nu)}_{\mathbf{N}_{\alpha}\mathbf{N}_{\alpha'}},
\label{eq:masterN}
\end{equation}
where $J^{(\nu)}_{\alpha\alpha'}(t)= \omega^{(\nu)}_{\mathbf{N}_{\alpha}\mathbf{N}_{\alpha'}}p_{\mathbf{N}_{\alpha'}}(t)- \omega^{(\nu)}_{\mathbf{N}_{\alpha'}\mathbf{N}_{\alpha}} p_{\mathbf{N}_{\alpha}}(t)$. 
The transition rates are written as 
\begin{equation}
\omega^{(\nu)}_{\mathbf{N}_\alpha\mathbf{N}_\alpha'} = \Gamma\, \Omega(N_{\alpha'})
\exp\left[-\frac{\beta_\nu}{2} Q^{(\nu)}_{\mathbf{N}_\alpha\mathbf{N}_{\alpha'}}\right],
\label{eq:rateN}
\end{equation}
where  $\Omega(N_{\alpha'}) = N_{\alpha'}$ denotes the degeneracy of the state associated with the occupation number of $\alpha'$ and $Q^{(\nu)}_{\mathbf{N}_\alpha\mathbf{N}_{\alpha'}}$ is the same as in Eq.~(\ref{heat1}) with 
the energy difference 
\begin{equation}
\Delta E_{\mathbf{N}_\alpha\mathbf{N}_{\alpha'}} =  \frac{\epsilon}{N}\left(N_\alpha - N_{\alpha'} + 1\right).
\label{eq:deltaEN}
\end{equation}

We now turn to the limit $N\rightarrow\infty$. In such case, the dynamics can be solely expressed in terms of the mean occupation density $n_\alpha\equiv\langle N_\alpha/N\rangle$ constrained by the normalization condition $\sum_\alpha n_\alpha = 1$. By employing a similar description to Refs.~\cite{herpich, herpich2, VANKAMPEN2007193}, each average of type $\langle \omega^{(\nu)}_{\alpha \alpha'}\, n_{\alpha'}\rangle$ can be rewritten as $\langle \omega^{(\nu)}_{\alpha\alpha'}\, N_{\alpha'}\rangle/N\approx \langle\omega^{(\nu)}_{\alpha\alpha'}\rangle \langle N_{\alpha'}\rangle/N\rightarrow \omega^{(\nu)}_{\alpha\alpha'} n_{\alpha'}$, where the energy difference appearing in transition rates reads $\Delta E_{\alpha\alpha'}=\epsilon(n_\alpha-n_{\alpha'})$. The time evolution in this limit is governed by the rate equation     
\begin{align}
    \dot{n}_{\alpha}(t)=\sum_{\nu=1}^2\sum_{\alpha'\neq\alpha}&
    J^{(\nu)}_{\alpha\alpha'}(t),\label{eq:master-equation}
\end{align}
where $J^{(\nu)}_{\alpha\alpha'}(t)=\omega^{(\nu)}_{\alpha\alpha'}n_{\alpha'}(t)-\omega^{(\nu)}_{\alpha'\alpha}n_{\alpha}(t)$.
In the nonequilibrium steady-state (NESS) regime, the densities are characterized by steady values ${n^{st}_{\alpha}}$. 
Our results for the mean-field model are obtained from the steady-state solution of these nonlinear equations.

\subsection{Two-dimensional model}

Except for the undriven case, corresponding to $F=0$, our model has no known exact analytical solution in two dimensions. For this reason, we investigate the 
phase transitions and evaluate the thermodynamic quantities for a square-lattice with periodic boundary conditions using the Gillespie algorithm \cite{gillespie1977exact}. 

A given microscopic configuration $s$ gives rise to a new configuration $s'$, 
associated with the $\nu$-th thermal reservoir, with probability $\omega^{(\nu)}_{s's}/W$, where $W=\sum_\nu\sum_{s'}\omega^{(\nu)}_{s's}$. The total rate $W$
accounts for the $2N(q-1)$ different configurations $s'$ that can be generated from $s$ through transitions mediated by either thermal reservoir. At each transition, the simulation time is incremented by $\Delta t=-\ln\xi/W$, where $\xi$ is a random number uniformly distributed between 0 and 1.
For the selected new configuration, the instantaneous exchanged heat and power are given by $Q^{(\nu)}_{s's}\equiv\Delta E_{s's}-F^{(\nu)}_{s's}$ and $F^{(\nu)}_{s's}$, respectively, where $\Delta E_{s's}=\epsilon\sum_{\delta=1}^{k}(\delta_{{\tilde s}_i,s_{i+\delta}}-\delta_{s_i,s_{i+\delta}})$. Additionally, the order-parameter quantities are evaluated from histograms based on the population of each state. 
Since Gillespie dynamics requires the generation of all $2N(q-1)$ possible configurations at each time step, it becomes computationally expensive as $N$ increases. For this reason, we restrict our analysis to system sizes ranging from $N=10^2$ to $20^2$. The steady-state densities $n^{\rm st}_\alpha$ are evaluated numerically using the Gillespie algorithm according to $n^{\rm st}_\alpha=\langle\sum_{i=1}^N\delta_{s_i,\alpha}\rangle/N$.

\subsection{Heat flux, power, and efficiency}

We now turn to the system thermodynamics. By taking the time evolution of the mean energy $ E\equiv N^{-1}\sum_s E_s p_s(t)$, we write the first-law of thermodynamics
\begin{equation}
\frac{dE}{dt}=Q_1+Q_2+P
\label{first}
\end{equation}(see e.g. Refs.~\cite{broeck15,gatien,forao2023powerful}), where $Q_\nu$ and $P$ denote the mean rate of absorbed heat per unit associated with the $\nu-$th thermal reservoir and the mean power per unit, respectively. The expressions of these quantities are given by
\begin{equation}
Q_\nu\equiv N^{-1}\sum_{(s,s'\neq s)}Q^{(\nu)}_{s's},{J}^{(\nu)}_{s's}\label{Heat_General}
\end{equation}
and
\begin{align}
    P\equiv N^{-1}F\sum_{\nu=1}^2&\sum_{s'<s}d^{(\nu)}_{ss'}J^{(\nu)}_{ss'}.
    \label{power2}
\end{align} 
The rate of entropy production per unit
is defined as \cite{schnakenberg1976network}
\begin{equation}
\sigma\equiv N^{-1}\sum_{\nu=1}^2\sum_{s,s'\neq s}J^{(\nu)}_{s',s}\ln\frac{\omega^{(\nu)}_{s',s}}{\omega^{(\nu)}_{s,s'}}=-\sum_{\nu=1}^2\beta_\nu Q_\nu\ge 0,
    \label{SS_General_Res}
\end{equation}
where the second equality is obtained from Eqs.~(\ref{transition}), (\ref{heat1}) and (\ref{Heat_General}).

The  expressions above are valid for both mean-field and two-dimensional models. However, in the former case we can express
heat and power in terms of the densities $n_\alpha$. These expressions are given by
\begin{equation}
Q_\nu=\sum_{\alpha'<\alpha}Q^{(\nu)}_{\alpha\alpha'}J^{(\nu)}_{\alpha\alpha'},
\label{heatdef}
\end{equation}
and 
\begin{equation}
P=F\sum_{\nu=1}^2\sum_{\alpha'<\alpha}d^{(\nu)}_{\alpha\alpha'}J^{(\nu)}_{\alpha\alpha'}.
\label{powdef}
\end{equation}

We are interested in a regime where the system operates as a heat-engine, converting heat absorbed from the thermal reservoirs into extracted power. The extracted power is $-P$, and the heat-engine efficiency is given by
\begin{equation}
\eta\equiv-\frac{P}{Q_2}.  
\end{equation}
From the first-law in Eq.~(\ref{first}) and second-law in Eq.~(\ref{SS_General_Res}), it follows that 
the efficiency $\eta$ is bounded by the Carnot efficiency, i.e., $\eta\le\eta_C\equiv1-\beta_2/\beta_1$. 
In our results, we also observe the system operating in two other regimes: a refrigerator, corresponding to $P>0$ and  $Q_2<0$, and a dud, corresponding to $P>0$ and  $Q_2>0$.

\section{Splitting of the ordered phase} \label{MF-results1}

{\subsection{Order Parameter}}

The undriven case, which remains out of equilibrium due to the contact with thermal reservoirs at different temperatures, has been considered in \cite{mamede2025exactmapping}. In this case, the order parameter $m$ is related to the densities $n_\alpha$ through the relations $n_{0}^{\rm st}=[1+(q-1)m]/q$ and ${n}^{\rm st}_{\alpha\neq 0}=(1-m)/q$. We observe that $m$ is invariant under the exchange $n_0\leftrightarrow n_\alpha$ ($\alpha \in \{1,...,q-1\}$), reflecting the $S_q$ permutation symmetry of the Potts model.


 
In contrast, in our model with $F\neq 0$, there is also a separation within the ordered phase. In particular, our scheme
for the driving breaks the $S_q$ symmetry by favoring transitions toward higher states for the hot reservoir and toward lower states for the cold reservoir. This splitting of the ordered phase is not captured by $m$ but rather by the observable 
\begin{equation}
\phi\equiv n^{\rm st}_0-n^{\rm st}_{q-1}.
\end{equation}
The quantity $\phi$ is a simple scalar that quantifies  the difference between the densities $n^{\rm st}_0$ and $n^{\rm st}_{q-1}$  within the ordered phase with $m\neq 0$. 
For differences within the intermediate values of $\alpha$, we have to look at the full phase-space of the densities $n_\alpha$.

\begin{figure*}
    \centering
    \includegraphics[scale=1]{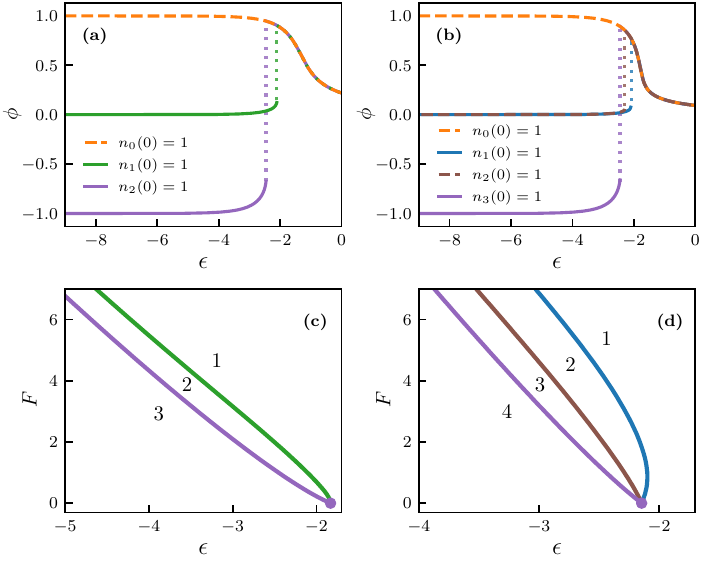}
    \caption{Main results of the phase transitions for MF interactions for $q=3$ and $q=4$. For $F=1$, panels $(a)$ and $(b)$ 
    depict the order parameter $\phi$ for $q=3$ and $q=4$, respectively, and different initial conditions $n_\alpha(0)$ ($\alpha \in \{0,..,q-1\}$). Panels $(c)$ and $(d)$ show the corresponding phase diagrams, where $1,2,3$ and $4$ denote the number of stable fixed points in that phase, and lines denote the discontinuous transitions. Parameters: $\beta_1=2$ and $\beta_2=1$.}
    \label{fig:transitions_MF}
\end{figure*}

\subsection{Phase transitions within the ordered phase}

The phase transitions within the ordered phase are depicted in Fig.~\ref{fig:transitions_MF} for the MF model for $q=3$ and $q=4$. 
We begin with $q=3$ in Fig.~\ref{fig:transitions_MF}(a), where we consider three different initial conditions: $n_{\alpha}(0)=1$ for $\alpha=0,1,2$. 
For large $-\epsilon$, there are three stable fixed points with $n_\alpha\approx 1$. This scenario is standard in the ordered phase of the Potts model.

\begin{figure*}[htb!]
    \centering
    \includegraphics[scale=1.]{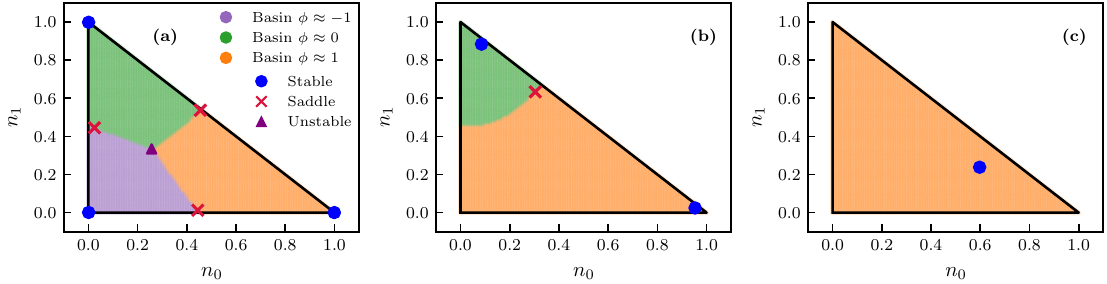}
    \caption{Basin of attraction for the fixed points for $q=3$. Each panel represent a phase, with (a) $\epsilon=-5$, (b) $\epsilon=-2.3$ and (c) $\epsilon=-1$. Parameters: $F=1$, $\beta_1=2$ and $\beta_2=1$.}
    \label{fig:basinq3}
\end{figure*}

As $-\epsilon$ decreases, the fixed points start to change from $n_\alpha\approx 1$, reducing the favoring for a state. At a certain critical value, the fixed point that favors $n_2$ disappears  and
there is a phase transition from three stable solutions to two stable solutions. As shown in Fig.~\ref{fig:basinq3}, the initial condition $n_{2}(0)=1$ becomes part of the basin of attraction 
of the stable fixed point dominated by $n_{0}$. By further decreasing  $-\epsilon$, at a second critical value, the fixed point dominated by $n_{1}$ 
(corresponding to $\phi=0$) also disappears and the stable fixed point dominated by $n_{0}$ is the only one that survives. 

Hence, for $q=3$, there are two discontinuous phase transitions within the ordered phase. For decreasing $-\epsilon$, in the first transition the system goes from three stable fixed points to two stable fixed points, 
and in the second transition it goes from two stable fixed points to one fixed point. A similar situation happens to $q=4$, shown in Fig.~\ref{fig:transitions_MF}(b), which has three phase transitions. 
For decreasing $-\epsilon$, first there are four stable fixed points, then the $n_3$ dominated solution disappears and there are three fixed points, 
then, after the second phase transition, the $n_2$ dominated solution disappears and there are two solutions, and, finally, after the third phase transition, the only fixed point is the one dominated 
by $n_0$. In all phase transitions, the fixed point that disappears becomes part of the basin of attraction of the fixed point dominated by $n_0$. 

The full phase diagrams in the $\epsilon\times F$ plane for $q=3$ and for $q=4$ are shown in Fig.~\ref{fig:transitions_MF}(c) and in Fig.~\ref{fig:transitions_MF}(d), respectively. The different 
phases are characterized by the number of stable fixed points. A similar situation should happen for higher values of $q$, with the ordered phase splitting into $q$ different phases characterized by the number of 
stable fixed points.
  
The parameter $\phi$ has the nice feature of being a simple scalar, however, it does not provide much information about the intermediate  densities $n_\alpha$. For a better picture, we look at the 
full phase space $n_1\times n_2$ for $q=3$ in Fig.~\ref{fig:basinq3}. In this case, the full phase space is two-dimensional since $n_0=1-n_1-n_2$. The picture shows that the basin of attraction of the fixed points 
depends on $\epsilon$. By decreasing $\epsilon$, the basin of attraction of the fixed point dominated by $n_0$ increases. After the first phase transition this basin of attraction becomes big enough 
to make the fixed point dominated by $n_2$ disappear, and after the second phase transition the fixed point dominated by $n_1$ is also caught by the basin of attraction of the fixed point dominated by
$n_0$.

\section{Emergence of a collective heat engine}\label{MF-results2}

\begin{figure*}[htb!]
    \centering
    \includegraphics[scale=1.]{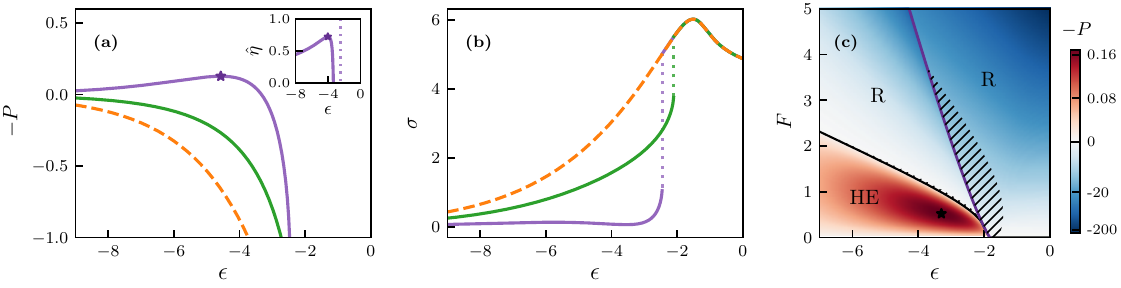}
    \caption{Main results of the thermodynamics for MF interactions for $q=3$. Panels $(a)$ and $(b)$ show the depiction
    of the mean power $-P$ and of the  entropy productions per unit $\sigma$, respectively, versus $\epsilon$ for $F=1$ 
    Inset: the scaled efficiency ${\hat \eta}=\eta/\eta_c$. Curves label are the same as in Fig.~\ref{fig:transitions_MF}. Panel $(c)$ show the power heat-map for $n_2(0)=1$ where $HE$ and $R$ denote, respectively, the heat-engine and the refrigerator regimes. The hatched region indicate the dud regime. The symbol $\star$ denotes the simultaneous maximization of $-P$
    with respect to $F$ and $\epsilon$. The purple continuous line indicate the phase transition as in Fig.~\ref{fig:transitions_MF}$(c)$.
    Parameters: $\beta_1=2$ and $\beta_2=1$. }
    \label{fig:thermodynamics_MF}
\end{figure*}

\subsection{Power, efficiency, and dissipation}

A remarkable feature of our model is that it can operate as a heat engine, as shown in Fig.~\ref{fig:thermodynamics_MF}. For $q=3$, the extracted power $-P$ is positive, corresponding to a heat engine, in  a region 
that has all three fixed points. The curve that has this heat engine regime is the one that corresponds to the fixed point dominated by $n_2$, as demonstrated in Fig.~\ref{fig:thermodynamics_MF}(a). 
Hence, the heat engine happens for initial conditions within the basin of attraction  of this fixed point. In other words, the interplay between the two temperatures and the driving associated with each temperature 
generates a heat engine in the phase with all $q$ fixed points and for initial conditions in the basin of attraction of the fixed point dominated by $n_{q-1}$, which is the first one to disappear 
for decreasing $-\epsilon$.

Interestingly, this heat engine regime for the fixed point dominated by $n_{q-1}$ corresponds to a much lower rate of entropy production $\sigma$ Fig.~\ref{fig:thermodynamics_MF}(b). Hence,
the heat engine regime is much less dissipative then the other two fixed points that cannot lead to power extraction. The phase diagram in Fig.~\ref{fig:thermodynamics_MF}$(c)$ shows 
the different regimes of operation: heat engine $HE$ and refrigerator $R$. There is also a region where the system behaves like a dud.

\begin{figure}[htb!]
    \centering
    \includegraphics[scale=1.]{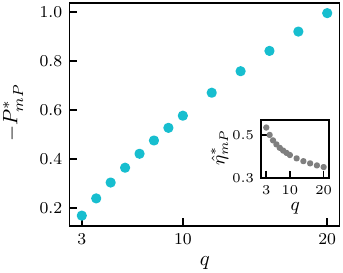}
    \caption{The dependence of maximal power $-P^{*}_{\text{mP}}$
    with $q$. Inset: The same but for $\hat{\eta}^{*}_{\text{mP}}$. Parameters: $\beta_1=2$ and $\beta_2=1$.}
    \label{fig:max}
\end{figure}

By simultaneously optimizing over $\epsilon$ and $F$, we obtain the global maximum power $-P_{\text{mP}}^{*}$ and its corresponding efficiency at 
maximum power, $\hat{\eta}_{\text{mP}}^{*} \equiv \eta(P_{\text{mP}}^{*})/\eta_{\text{C}}$. In Fig. \ref{fig:max}, we show how both quantities depend on the 
number of states $q$. The global maximum power $-P^{*}_{\text{mP}}$ grows with $q$ and the corresponding efficiency at maximum power decreases with $q$.

\subsection{Analytical approximation for the MF model}

To gain analytical insight into the heat engine performance in the mean-field model, we introduce a phenomenological linearization of the rate equations in the strongly ordered regime. 
For the phase with $q$ fixed points and for the fixed point dominated by the state $q-1$, we assume that $n_{q-1}^{\text{st}} \approx 1$, which should be the case for $-\beta_\nu\epsilon\gg 1$. We also assume 
that the other states are sparsely and symmetrically populated, i.e.,
\begin{equation}
    n_{\alpha}^{\text{st}} \approx \frac{1 - n_{q-1}^{\text{st}}}{q-1} \ll 1 \quad \text{for all } \alpha \neq q-1.
\end{equation}

In this limit, transitions among the minority states $\alpha, \alpha' \neq q-1$ contribute negligibly to the overall dynamics ($J_{\alpha\alpha'}^{(v)} \approx 0$). 
The rate equations, given in Eq. \eqref{eq:master-equation}, simplify to a linear star-like network where the dominant state $q-1$ exchanges density independently with each of the other $q-1$ states with $\alpha\neq q-1$. 
The steady state solution of this linearized set of equations reads 
\begin{equation}
    n_{q-1}^{\text{eff}} = \frac{1}{1 + (q-1) e^{\frac{1}{2}\left[\epsilon(\beta_1 + \beta_2) + F(\beta_1 - \beta_2)\right]}} = \frac{1}{1 + (q-1) \omega_1 \omega_2},
 \label{eq18}
\end{equation}
where the effective single-step forward and backward transition rates are defined as $\omega_1 \equiv e^{\frac{1}{2}\beta_1(\epsilon + F)}$ and $\omega_2 \equiv e^{\frac{1}{2}\beta_2(\epsilon - F)}$.

Because the linearized topology lacks cycles among minority states, detailed balance holds across each individual dominant-to-minority transition pathway. The total net probability current driven
 across the $q-1$ equivalent transitions is described by a single effective current
\begin{equation}
    J_{\text{eff}} \equiv (q-1) J_{q-1, \alpha}^{(1)} = \Gamma (q-1) (\omega_2 - \omega_1) n_{q-1}^{\text{eff}}.
\end{equation}
Using $J_{\text{eff}}$, the thermodynamic quantities adopt simple, closed-form analytical expressions:
\begin{align}
    Q_{1, \text{eff}} &= (\epsilon + F) J_{\text{eff}}, \\
    Q_{2, \text{eff}} &= (F - \epsilon) J_{\text{eff}}, \\
    P_{\text{eff}} &= -2F J_{\text{eff}}, \\
    \eta_{\text{eff}} &= -\frac{P_{\text{eff}}}{Q_{2, \text{eff}}} = \frac{2F}{F - \epsilon}.
\end{align}
Therefore, within this approximation there is tight coupling with all thermodynamic fluxes proportional to a single current $J_{\text{eff}}$.
Setting $\omega_1 = \omega_2$ yields $J_{\text{eff}} = 0$, which defines the stalling condition where extracted power vanishes and the system would achieve Carnot efficiency.

\begin{figure}[htb!]
    \centering
    \includegraphics[scale=1.]{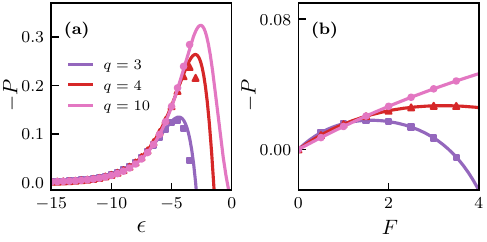}
    \caption{The depiction of power and the comparison between exact 
    $-P$ (symbols) and phenomenological $-P_{\rm eff}$ (continuous lines) descriptions. Parameters: In all cases we set $\beta_1=2,\beta_2=1$, $F=1$~panel~$(a)$ and $\epsilon=-10$ panel~$(b)$.}
    \label{fig:approx}
\end{figure}

In Fig.~\ref{fig:approx} we compare  analytical results, with this approximation, with exact results for the nonlinear rate equation. They show excellent agreement
for larger $-\beta_\nu\epsilon$'s and the appearance of discrepancies as $-\beta_\nu\epsilon$ is lowered. 

For $-\epsilon$ large such that $\omega_1\omega_2<<1/(q-1)$, the denominator of Eq.~\eqref{eq18} can be
neglected and
$-P_{\rm eff}\simeq 2\Gamma F(q-1)(\omega_2-\omega_1)$. Since $\omega_1$
and $\omega_2$ do not depend on $q$, the number of states then enters
only through the overall factor $q-1$: it sets the magnitude of the
extracted power, but not the location of its optimum. Accordingly, the
maximization of $-P_{\rm eff}$ with respect to $\epsilon$ ($F$ fixed)
and $F$ ($\epsilon$ fixed) yields the $q$-independent conditions
\begin{equation}
\frac{\omega_2}{\omega_1}=\frac{\beta_1}{\beta_2}
\qquad\text{and}\qquad
\frac{\omega_2}{\omega_1}=\frac{2+\beta_1F_{\text{mP}}}{2-\beta_2F_{\text{mP}}},
\label{opt}
\end{equation}
respectively. The first one is explicit,
\begin{equation}
\epsilon_{\text{mP}}=-\frac{(\beta_1+\beta_2)F+2\ln(\beta_1/\beta_2)}{\beta_1-\beta_2},
\label{epsmP}
\end{equation}
whereas the second restricts the useful driving to $F_{\text{mP}}<2/\beta_2$.

Maximizing the analytical power $-P_{\text{eff}}$ with respect to the driving force $F$ and coupling parameter $\epsilon$ leads to the optimal parameters at maximum power:
\begin{eqnarray}
F_{\rm mP}&=&\frac{1}{\beta_2}-\frac{1}{\beta_1},\label{eq:FmP}\\
\epsilon_{\rm mP}&=&-\left(\frac{1}{\beta_1}+\frac{1}{\beta_2}\right)
-\frac{2\ln(\beta_1/\beta_2)}{\beta_1-\beta_2}.\label{eq:epsmP}
\end{eqnarray}
Because the number of states $q$ only enters $J_{\text{eff}}$ as an overall scaling factor $(q-1)$, the optimal parameters $F_{\text{mP}}$ and $\epsilon_{\text{mP}}$ 
are completely independent of $q$ with this approximation valid in the strongly ordered regime.

While the phenomenological model accurately captures the scaling behavior in the strongly ordered limit ($-\beta_v \epsilon \gg 1$), 
the simultaneous optimization with respect to both $\epsilon$ and $F$ shifts the optimal operation point to a regime with smaller $-\epsilon$. 
In this regime, minority-state population and cross-pathway transitions cannot be neglected, necessitating the full numerical optimization shown in 
Fig. \ref{fig:thermodynamics_MF}(c) and Fig. \ref{fig:max}.

\section{Results for the Two-dimensional model} \label{SL-results}

\subsection{Phase transition }

\begin{figure}
    \centering
    \includegraphics[scale=1.]{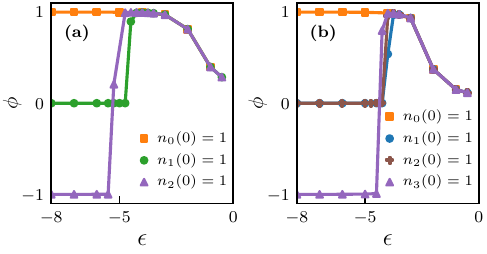}
    \caption{For the same parameters as in Fig.~\ref{fig:transitions_MF}, the depiction
    of the order-parameter $\phi$ for $q=3$ (left) and $q=4$ (right) for square-lattice topologies. Lines connect the points for visual guidance.}
    \label{fig22}
\end{figure}

Fig.~\ref{fig22} depicts the results for the order parameter for $q=3$ and $q=4$ using local interactions. We observe the same qualitative behavior of the 
MF model in the thermodynamic limit, demonstrating the robustness of the phenomenology with respect to the topology of the lattice and to finite size effects. 
The real stationary state of the finite stochastic system should be independent of the initial conditions. 
In fact, although the results show a ``metastable'' state for a finite system, the time to relax to the stationary state 
grows exponentially with  systems size and becomes prohibitively large.  We observe that for  rather long observation 
times of $5.10^6$ and  system sizes $N\le 400$ we do not see the onset of the stationary state. 

In conclusion, we demonstrated that the phenomena of splitting of the ordered phase characterized by the  number of stable solutions,
is observed in a finite system with finite dimension. The multiple stationary solutions of the MF model in the thermodynamic limit 
become long-lived metastable states for finite lattices. It is worth pointing out that transition points  are different from the ones in the MF model, and they are closer to each other in comparison.
 Numerical simulations also indicate that phase transitions are discontinuous, similar to MF. 

\subsection{Thermodynamics}

\begin{figure}
    \centering
    \includegraphics[scale=1.]{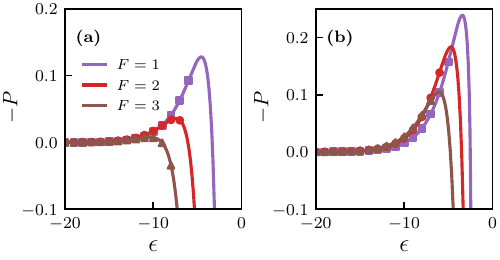}
    \caption{The depiction of $P$ 
    for square-lattice (symbols) versus
    $\epsilon$ for different
    $F$'s and $q=3$~(left) and $q=4$~(right). For the sake of comparison, results for MF case
    (continuous lines) are shown as well. Parameters: $\beta_1=2,\beta_2=1$ and system size $N=100$. }
        \label{fig:power_SL}
\end{figure}
\begin{figure}
    \centering
    \includegraphics[scale=1.]{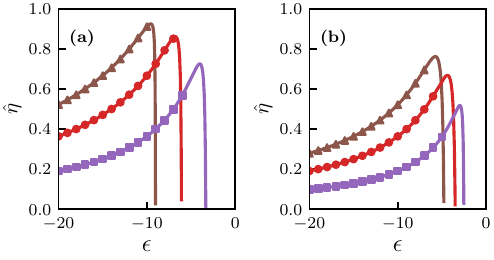}
    \caption{The same as in Fig.~\ref{fig:power_SL} but
    for the scaled efficiency $\hat \eta$. }
        \label{fig:eff_SL}
\end{figure}

The two-dimensional model can also behave as a heat engine in the phase with all $q$ solutions and for the solution dominated by $n_{q-1}$. This fact is demonstrated by 
plotting the extracted power $-P$ in Fig.~\ref{fig:power_SL} and the rescaled efficiency $\hat{\eta}$ in Fig. \ref{fig:eff_SL}. 
For the parameters we have observed, when the model behaves as a heat engine, the interaction $\epsilon$ is large enough so that power and efficiency for MF and two-dimensional models 
seem to agree with each other, as shown in Figs.~\ref{fig:power_SL} and \ref{fig:eff_SL}.

\section{Conclusions} \label{conclusion}                 

For the present driven Potts model in contact with two heat baths, we have demonstrated the emergence of nonequilibrium phase transitions, within the ordered phase. 
They arise from the interplay between coupling to thermal reservoirs at different temperatures and 
nonconservative driving forces. The different phases are characterized by the different number of 
stable fixed points, with $q-1$ transitions. From the typical phase with $q$ stable fixed points to a phase with one fixed point
dominated by the density $n_0$.  

The second remarkable feature of our model is the emergence of a collective heat engine. 
This heat engine is realized within the phase where all $q$ stable solutions are present and 
for initial conditions within the basin of attraction of the stable fixed point dominated by the density $n_{q-1}$. Therefore,
the interplay between the two heat reservoirs and the driving force $F$ generates two a priori separate phenomena. The splitting 
of the ordered phase into $q$ different phases and  the emergence of a collective heat engine. Despite the complexity of the 
underlying dynamics and thermodynamics, we developed a phenomenological description that captures the main features of the system for the MF model that allows 
for closed-form expressions for heat flux and power..

These phenomena are robust with respect to dimensionality and for finite systems with fluctuations.  We have observed that the two-dimensional model displays both features, 
the splitting of the ordered phase into $q$ phases and the emergence of a heat engine. Due to the lack of an analytical solution we resorted 
to numerical simulations of finite systems for the two-dimensional model.  

As an interesting direction for future work, it remains an open question whether the splitting of the ordered phase and the emergence of a heat engine 
are phenomena that are coupled together. Here, we achieved these two phenomena by using a force that breaks the symmetry between states of the Potts model. 
The minimal ingredients for these two features remain as an open question.


\section{Acknowledgments}
VTH, GALF and CEF acknowledge the financial support from FAPESP under grants 2024/08157-0,  2022/15453-0, 2022/16192-5, 2024/03763-0 and
2023/17704-2. The financial support from CNPq is also acknowledged. ACB acknowledges the financial support from the NSF
through the grant DMR-2424140.

\bibliography{bib}

\end{document}